\documentclass{article}
\usepackage{icmc2025_paper_template}
\usepackage{times}
\usepackage{ifpdf}
\usepackage{soul}
\usepackage[english]{babel}
\usepackage[numbers]{natbib} 
\def\papertitle{Dialogue in Resonance: An Interactive Music Piece for Piano and Real-Time Automatic Transcription System}

\def\firstauthor{First Author}
\def\secondauthor{Second Author}
\def\thirdauthor{Third Author}

\newif\ifpdf
\ifx\pdfoutput\relax
\else
   \ifcase\pdfoutput
      \pdffalse
   \else
      \pdftrue
  \fi
\fi

\ifpdf 
  \usepackage[pdftex,
    pdftitle={\papertitle},
    pdfauthor={\firstauthor, \secondauthor, \thirdauthor},
    bookmarksnumbered, 
    pdfstartview=XYZ 
   ]{hyperref}

  \usepackage[pdftex]{graphicx}
  \graphicspath{{./figures/}}
  \DeclareGraphicsExtensions{.pdf,.jpeg,.png}

  \usepackage[figure,table]{hypcap}

\else 
  \usepackage[dvips,
    bookmarksnumbered, 
    pdfstartview=XYZ 
  ]{hyperref}  

  \usepackage[dvips]{epsfig,graphicx}
  \graphicspath{{./figures/}}
  \DeclareGraphicsExtensions{.eps}

  \usepackage[figure,table]{hypcap}
\fi

\hypersetup{
    colorlinks,%
    citecolor=black,%
    filecolor=black,%
    linkcolor=black,%
    urlcolor=black
}

\title{\papertitle}

\author{
   \textbf{Hayeon Bang} \hspace{4em} 
   \textbf{Taegyun Kwon} \hspace{4em}
   \textbf{Juhan Nam} \\[0.5em]
   Graduate School of Culture Technology, KAIST\\ 
   {\tt \{hayeonbang, ilcobo2, juhan.nam\}@kaist.ac.kr}
}

\begin{document}
\capstartfalse
\maketitle
\capstarttrue
\begin{abstract}
This paper presents \textit{Dialogue in Resonance}, an interactive music piece for a human pianist and a computer-controlled piano that integrates real-time automatic music transcription into a score-driven framework. Unlike previous approaches that primarily focus on improvisation-based interactions, our work establishes a balanced framework that combines composed structure with dynamic interaction.\\ Through real-time automatic transcription as its core mechanism, the computer interprets and responds to the human performer's input in real time, creating a musical dialogue that balances compositional intent with live interaction while incorporating elements of unpredictability. In this paper, we present the development process from composition to premiere performance, including technical implementation, rehearsal process, and performance considerations.
\end{abstract}

\section{Introduction}\label{sec:introduction}

The evolution of technology has continuously developed the ways of composing, performing, and experiencing music. From audio recording technology and electronic instruments to algorithmic music composition and AI-driven music generation, each technological advancement has contributed to expanding the boundaries of music. These advancements have offered artists new tools for experimentation in composition and performance. As these technologies evolved, the role of computers in music creation has transformed from passive tools to active participants, leading to the emergence of interactive music systems.

Interactive music systems, which respond to musical inputs from humans and participate in live performances, have evolved through several distinct phases \cite{rowe}. The seminal work of Rowe et al. established foundational concepts for these systems and categorized them according to various perspectives, while \textit{Synthetic Performer} by Vercoe et al. \cite{inter_synthetic} and \textit{Duet for One Pianist} by Risset et al. \cite{risset} demonstrated practical implementations of interactive music. Later, Lewis et al. \cite{lewis} and Dannenberg et al. \cite{im_bob} presented a system where a computer functioned as an autonomous improviser.

Recent studies have explored a broader range of computational techniques for interactive music, such as real-time tracking, algorithmic transformations, and machine learning-driven interaction. For instance, a virtual orchestral system was developed using rule-based synchronization \cite{inter_orchestra}. Others explored an automatic accompaniment system with basic improvisation techniques for duet interaction by learning from human duet performances \cite{im_xia2017}, while Lin et al. \cite{inter_human} introduced a human-computer duet system that synchronizes a virtual performer with a live musician.

Building on these advancements, \textit{Dialogue in Resonance} explores how interactive music systems can facilitate a musical dialogue between a human and a computer. The piece reconstructs the piano duet as a dynamic interaction, where the computer responds in real time through reproduction, transformation, and reinterpretation of the human performer’s input. Rather than positioning the computer as a passive tool or an autonomous generator, this work establishes it as an active collaborator in a genuine musical dialogue, challenging traditional notions of human-computer interaction in musical performance.

In the piece \textit{Dialogue in Resonance}, the integration of real-time automatic music transcription with a structured compositional framework creates a unique balance between control and spontaneity. While previous attempts at using real-time transcription in interactive music have primarily focused on improvisation-based interactions \cite{im_bob, im_xia2017, guitar, im_jazz}, our approach maintains a structured musical framework. To the best of our knowledge, this is the first attempt to create a score-driven interactive music piece that incorporates real-time transcription as a core mechanism for musical interaction, embracing both the unpredictability of live transcription and the intentionality of composed structure. This approach demonstrates how computational advancements can serve artistic vision, creating a new paradigm for human-computer musical dialogue that remains true to the composer's creative intent while embracing the dynamic possibilities of real-time interaction.

\textit{Dialogue in Resonance} has been premiered in September 2024.
This paper presents a comprehensive documentation of our journey from composition to performance, including the compositional process incorporating real-time music transcription, technical implementation using Max/MSP, and the rehearsal and performance processes. Through this paper, we share insights into the challenges and possibilities of creating interactive music that bridges traditional composition with real-time automatic music transcription. A performance video, along with the score and system demonstration, is accessible at the web demo\footnote{\url{https://hayeonbang.github.io/Dialogue_in_Resonance/}}. 

\section{Related Work}\label{sec:related_works}

\begin{figure*}[h]
\centering
\includegraphics[width=0.98\textwidth]{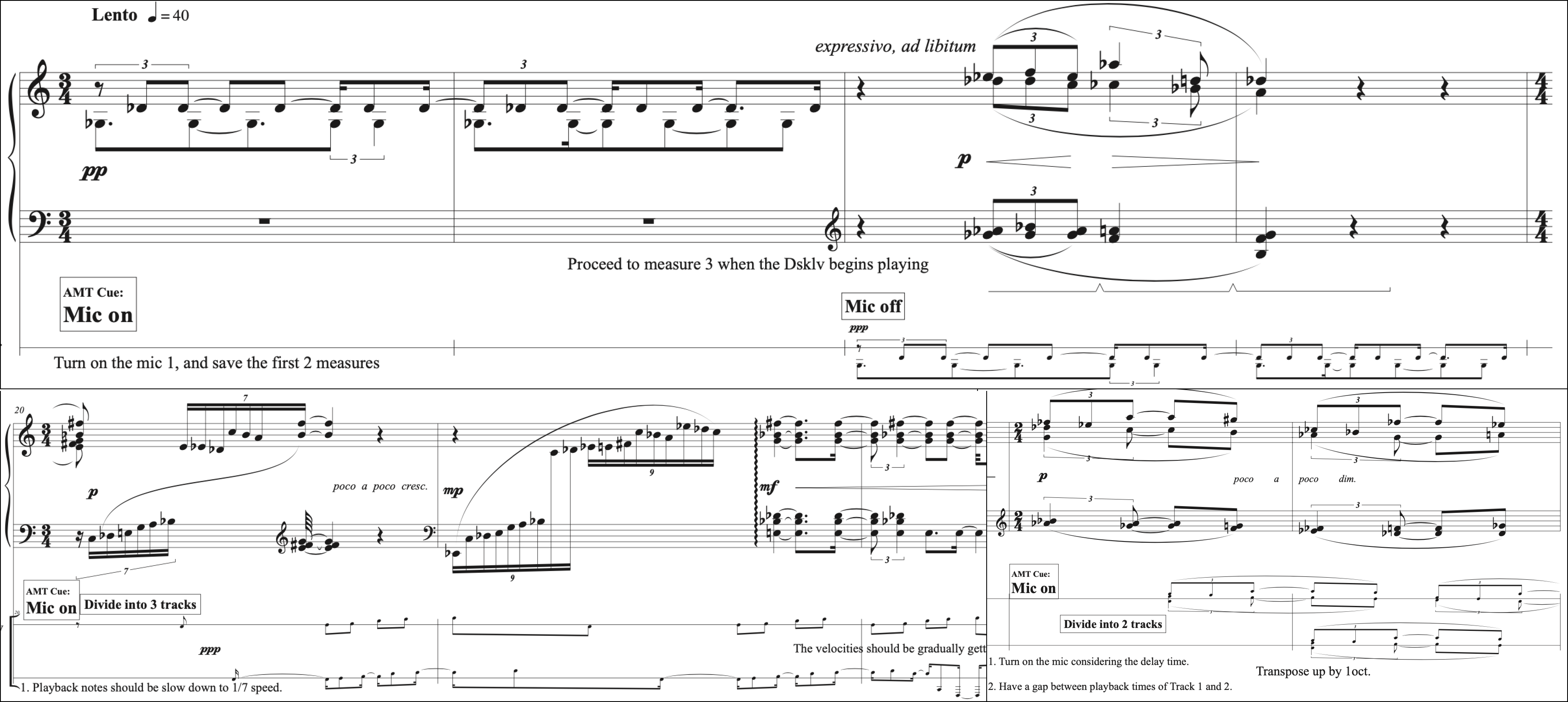}
\caption{Score excerpt from \textit{Dialogue in Resonance}. The first staff represents Piano 1, and the second staff represents Piano 2.\label{fig:score}}
\end{figure*}
Following the conceptual framework presented in Rowe et al. \cite{rowe},
interactive music systems can be categorized into score-driven systems (based on predefined scores) and performance-driven systems (based on real-time improvisational performance). They also defined response types such as Transformative, Generative, and Sequenced, providing a systematic understanding of how these systems function. This foundational framework has been influential in shaping subsequent research. Our work builds upon integrating score-driven elements, where both the human performer and the computer-controlled piano perform based on predefined scores and dynamics, and performance-driven elements, where the music incorporates real-time transcription of the human pianist’s performance, allowing for spontaneous and unpredictable notes. This duality creates a system that merges structured composition with dynamic interactivity.

Similar to our work, Risset et al. \cite{risset} presented a piece \textit{Duet for One Pianist}, which laid the groundwork for integrating MIDI technology to enable computer control of acoustic pianos through the use of eight distinct sketches. Their exploration of real-time interaction using the Yamaha Disklavier involved analyzing MIDI data from the performer to generate reflective or transformative responses, significantly expanding the expressive possibilities of the piano through technological augmentation. Building on this approach, Kallionpää et al. \cite{dis_game} developed \textit{Climb!}, a game-inspired interactive performance system for Disklavier, where the composition unfolds dynamically based on the performer’s musical choices. This work integrates real-time score-driven interactivity, enabling performers to shape the overall structure of the piece through specific musical triggers.

Lewis et al. \cite{lewis} introduced a piece \textit{Voyager}, redefining interactive music systems by demonstrating how computers could function as independent and equal musical agents. The system analyzed input from human performers to generate improvised responses while also producing autonomous musical content based on its internal algorithms. This non-hierarchical, subject-to-subject interaction emphasized unpredictability and improvisation in human-computer musical dialogue. 

More recently, Freire et al. \cite{guitar} introduced an interactive system based on automatic music transcription, similar to the approach we utilized. Their system used a hexaphonic guitar to analyze real-time transcription data and generate transformed musical outputs. By combining the traditional characteristics of the guitar with digital augmentation, this system enabled performers to manipulate or repeat specific musical segments interactively.

These prior studies have independently explored real-time responses, improvisation, and transcription-based interaction, collectively pushing the boundaries of interactive music systems. Our work synthesizes these elements, redefining the relationship between the computer-controlled piano and the human pianist. This study proposes an approach to interactive music that balances predictability with unpredictability, opening up new possibilities for creative expression.


\section{Methodology}
\subsection{Real-time Automatic Music Transcription}
In this paper, Automatic Music Transcription (AMT) refers to the process of converting the audio signal of an instrument into MIDI notes that reflect both the timing and velocity of the performance. Conventional AMT systems typically operate in an offline manner, meaning that transcription occurs only after the recording has been completed. Consequently, real-time interaction between performance and transcription is not possible in such systems.

To enable simultaneous performance and transcription, we utilize an online AMT approach, where the transcribed notes are generated in real time and can be used immediately as a musical response. Specifically, we employ the algorithm proposed by Kwon et al. \cite{amt}. This algorithm leverages deep neural networks to transcribe piano audio into MIDI signals with an approximate latency of 350 ms.

While the transcription accuracy depends on factors such as pitch and recording conditions, the original paper reports that the algorithm can achieve an onset F1 score of over 95\% in ideal conditions across various datasets.

In our system, the transcribed MIDI signals are not sent directly to the Disklavier. Instead, they are processed through Max/MSP, where delays and transformations are applied before the final performance, as detailed in Section \ref{subsec:implementation}.

\subsection{Compositional Process}
The piece \textit{Dialogue in Resonance} is designed with two pianos standing side by side, engaging in an interactive musical dialogue. The initial motivation for this work was to create a composition that showcases the musical communion between a human pianist (the first piano) and a computer-controlled piano (the second piano). To achieve this, we adopted a two-piano configuration, and to maximize musical interaction between the pianos, we implemented real-time transcription as the primary method.

The second piano does not perform pre-composed material; instead, it “listens” to the performance of the human pianist by analyzing, interpreting, and responding to it in real time. Also, it is giving responses by selectively reproducing, transforming, or even distorting the material it has perceived. This interactive mechanism allows the second piano to act as a responsive and interpretive partner rather than a passive executor of pre-composed content. 

To maintain a coherent musical framework, the first piano follows a score-driven structure, providing stability and continuity throughout the performance. Meanwhile, the second piano’s responses are guided by predetermined dynamics and intended musical expressions.

Although this piece is composed within a structured framework, it intrinsically incorporates elements of unpredictability. Each performance varies based on numerous live factors, such as the subtle shifts in the microphone activation timing, the resonance of the performance space, the human pianist’s interpretive choices, and the accuracy of the transcription process. The second piano does not merely replicate or reinterpret the first piano’s input; instead, it actively engages in collaborative interplay, complementing and responding to the human pianist’s performance. Through this interaction, the second piano becomes a vessel for reshaping the temporal and spatial dimensions of the music, creating a dialogue that evolves in real time. 

To musically reflect this sense of mystery and transformation, the composition employs tonal structures and harmonies imbued with a mystical character. These tonal colors emphasize the enigmatic and resonant qualities of the dialogue, making the interaction between the two pianos not only a technical-experimental process but also a musical piece. Figure \ref{fig:score} is a score excerpt from \textit{Dialogue in Resonance}.

\subsection{System Implementation}
\label{subsec:implementation}
\begin{figure}[ht]
\centering
\includegraphics[width=0.8\columnwidth]{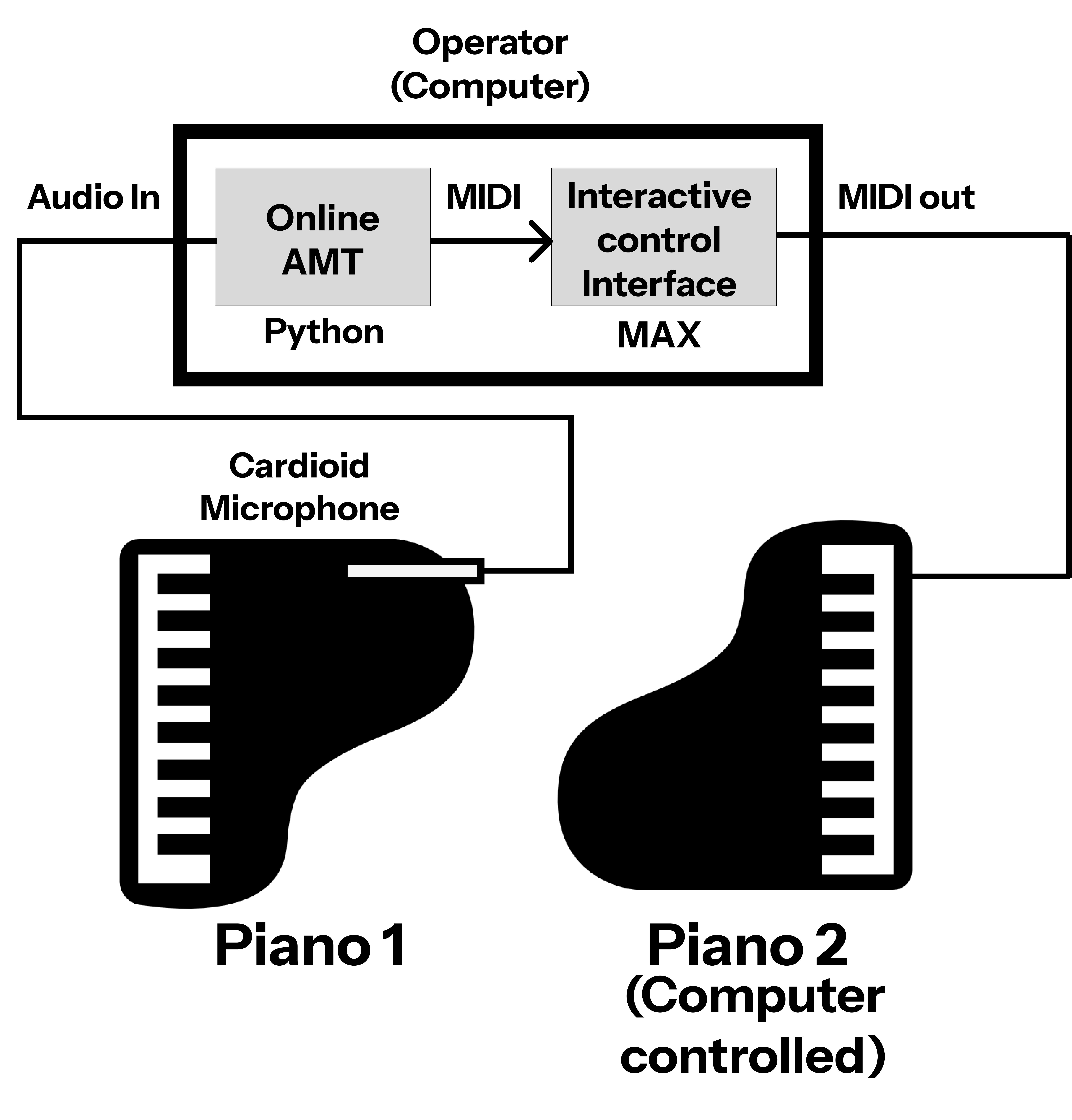}
\caption{Stage and system setup of \textit{Dialogue in Resonance}.\label{fig:stage}}
\vspace{-2mm}

\end{figure}
Since different sections of the piece required distinct effects using the same AMT system, we needed to dynamically adjust system parameters throughout the performance. Additionally, fine control was necessary to synchronize the computer-controlled piano's responses with the live acoustic environment. Rather than fully automating these adjustments, we designed an interface that allows a computer operator to fine-tune the system in real time.

An overview of the stage and the setup of the system is presented in Figure~\ref{fig:stage}. Audio from the first piano is captured using a cardioid microphone to suppress the sound from the second piano. The online AMT model processes the captured audio and converts it into MIDI notes in real time. These MIDI notes are then be transformed via an interactive control interface in Max, which modifies timing, dynamics, and playback behaviors before sending the final MIDI data to the computer-controlled second piano.

The system architecture consists of two main components:
\begin{itemize}
    \item \textbf{Python}, which runs the \textbf{online AMT} and transmits the transcribed MIDI data to \textbf{Max} via OSC (Open Sound Control).
    \item \textbf{Max/MSP}, which provides both \textbf{MIDI processing functions} and an \textbf{interactive control interface} for real-time adjustments.
\end{itemize}

\subsubsection{Features Implemented in Max}
The following functions were implemented in Max to manipulate the MIDI output:


\begin{enumerate}
    \item \textbf{Looping and Repeating MIDI Segments} – Allows recorded MIDI phrases to be looped and replayed.
    \vspace{-1.5mm}

    \item \textbf{Delayed MIDI Playback} – Enables MIDI notes to be played back with a controlled delay.
    \vspace{-1.5mm}

    \item \textbf{Variable Playback Speed} – Adjusts the speed of the MIDI playback for expressive timing control.
    \vspace{-1.5mm}

    \item \textbf{Dynamic Adjustment} – Modifies the velocity of the MIDI notes to control dynamics.
    \vspace{-1.5mm}

    \item \textbf{Sustain Pedal Control} – Adjusts the sustain pedal for more natural phrasing.
    \vspace{-1.5mm}

    \item \textbf{Stopping MIDI Playback} – Allows for immediate halting of MIDI playback when needed.
\end{enumerate}

These controls were implemented using buttons and sliders in Max, allowing the operator to interactively modify parameters. Since multiple functions could be applied simultaneously, certain sections of the performance utilized a combination of effects. For example, to play the lower-right section of Figure~\ref{fig:score}, both MIDI looping and delayed playback were applied together to create multiple layers of resonance.

\vspace{-2mm}

\section{Performance Preparation and Execution}

\begin{figure}
\centering
\includegraphics[width=\columnwidth]{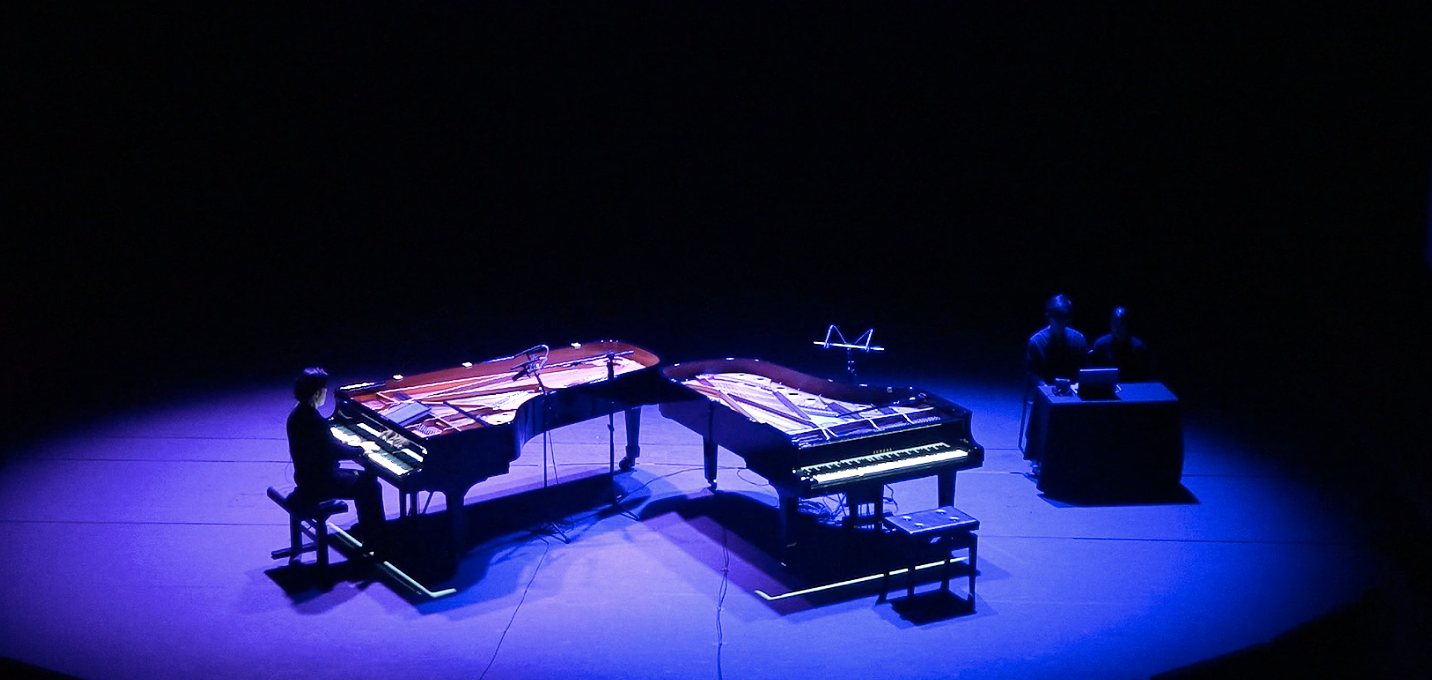}
\caption{Stage image of the live performance
\label{fig:perform}}
\vspace{-5mm}

\end{figure}

\textit{Dialogue in Resonance} premiered in September 2024 at the Ensemble Hall of the Daejeon Arts Center, Korea, as part of the \textit{Performance Laboratory: X-Space}. 
During preliminary rehearsals in the small studio, the transcription system occasionally struggled to distinguish subtle nuances in the first piano’s performance due to the less controlled acoustic conditions of the practice room. However, during stage rehearsals, the sound was significantly clearer and more direct, allowing the transcription to operate with much greater accuracy. For the performance, additional microphones were installed on each piano to ensure proper sound projection throughout the hall. As a result, the transcription microphone was repositioned to the left side of the first piano to accommodate the new stage setup and to maintain optimal capture of the pianist’s performance. 

The staging was designed to reflect the intention of the piece; the performance began in darkness, allowing the interplay of sound between the two pianos to take center stage, free from visual cues. This opening effect subtly blurred the perception of where the sound originated, setting an intriguing tone for the rest of the piece. Yamaha C3X Disklavier was used for the second piano, and an operator managed the computer system, with a backup operator.

To assess the system’s artistic and practical impact, we interviewed two pianists and two non-author operators. One pianist described a deep sense of immersion and co-creation, calling the second piano “another self.” An operator noted the system’s complexity but emphasized its expressive potential when well synchronized with the performer. Both groups recognized unpredictability as a valuable creative feature and stressed the importance of rehearsal and communication. Complete transcripts of these interviews with two pianists and two operators are available on the demo page\footnotemark[1].

\section{Conclusion}
The piece \textit{Dialogue in Resonance} demonstrates a practical approach to integrating real-time music transcription into score-driven interactive music. Through the premiere performance and rehearsal process, we gained valuable insights into both technical and artistic considerations of this approach. The implementation challenges, particularly regarding acoustic conditions and microphone placement, provided important lessons for future performances. Our experience suggests that real-time transcription can effectively serve compositional intent while maintaining the dynamic nature of live interaction. This work contributes to the field by providing a practical framework for creating interactive music that balances structure and spontaneity.
\begin{acknowledgments}
This work was supported by the collaboration with the Daejeon Arts Center, Korea, and funded under Grants RS-2023-NR077289 and RS-2024-00358448.
\end{acknowledgments} 

\bibliography{icmc2025_paper_template}
\end{document}